\documentstyle[12pt,aaspp4,epsfig]{article}
\def\spose#1{\hbox to 0pt{#1\hss}}
\newcommand\lsim{\mathrel{\spose{\lower 3pt\hbox{$\mathchar"218$}}
     \raise 2.0pt\hbox{$\mathchar"13C$}}}
\newcommand\gsim{\mathrel{\spose{\lower 3pt\hbox{$\mathchar"218$}}
     \raise 2.0pt\hbox{$\mathchar"13E$}}}

\def\ltsima{$\; \buildrel < \over \sim \;$}
\def\simlt{\lower.5ex\hbox{\ltsima}} % < over ~
\def\gtsima{$\; \buildrel > \over \sim \;$}
\def\simgt{\lower.5ex\hbox{\gtsima}} % > over ~

\def\V606{\hbox{$\rm V_{606}$}}
\def\I814{\hbox{$\rm I_{814}$}}
\def\I{\hbox{\rm I}}
\def\V{\hbox{$\rm V$}}

\newcommand{\etal} {{\it et~al.\ }}

\makeatletter
\let\@internalcite\cite
\def\cite{\def\@citeseppen{-1000}%
    \def\@cite##1##2{(##1\if@tempswa , ##2\fi)}%
    \def\citeauthoryear##1##2##3{##1 ##3}\@internalcite}

\def\citeNP{\def\@citeseppen{-1000}%
    \def\@cite##1##2{##1\if@tempswa , ##2\fi}%
    \def\citeauthoryear##1##2##3{##1 ##3}\@internalcite}
\def\citeN{\def\@citeseppen{-1000}%
    \def\@cite##1##2{##1\if@tempswa , ##2)\else{)}\fi}%
    \def\citeauthoryear##1##2##3{##1 (##3}\@citedata}
\def\citeA{\def\@citeseppen{-1000}%
    \def\@cite##1##2{(##1\if@tempswa , ##2\fi)}%
    \def\citeauthoryear##1##2##3{##1}\@internalcite}
\def\citeANP{\def\@citeseppen{-1000}%
    \def\@cite##1##2{##1\if@tempswa , ##2\fi}%
    \def\citeauthoryear##1##2##3{##1}\@internalcite}
\def\shortcite{\def\@citeseppen{-1000}%
    \def\@cite##1##2{(##1\if@tempswa , ##2\fi)}%
    \def\citeauthoryear##1##2##3{##2 ##3}\@internalcite}
\def\shortciteNP{\def\@citeseppen{-1000}%
    \def\@cite##1##2{##1\if@tempswa , ##2\fi}%
    \def\citeauthoryear##1##2##3{##2 ##3}\@internalcite}
\def\shortciteN{\def\@citeseppen{-1000}%
    \def\@cite##1##2{##1\if@tempswa , ##2)\else{)}\fi}%
    \def\citeauthoryear##1##2##3{##2 (##3}\@citedata}
\def\shortciteA{\def\@citeseppen{-1000}%
    \def\@cite##1##2{(##1\if@tempswa , ##2\fi)}%
    \def\citeauthoryear##1##2##3{##2}\@internalcite}
\def\shortciteANP{\def\@citeseppen{-1000}%
    \def\@cite##1##2{##1\if@tempswa , ##2\fi}%
    \def\citeauthoryear##1##2##3{##2}\@internalcite}
\def\citeyear{\def\@citeseppen{-1000}%
    \def\@cite##1##2{(##1\if@tempswa , ##2\fi)}%
    \def\citeauthoryear##1##2##3{##3}\@citedata}
\def\citeyearNP{\def\@citeseppen{-1000}%
    \def\@cite##1##2{##1\if@tempswa , ##2\fi}%
    \def\citeauthoryear##1##2##3{##3}\@citedata}

\def\@citedata{%
        \@ifnextchar [{\@tempswatrue\@citedatax}%
                                  {\@tempswafalse\@citedatax[]}%
}

\def\@citedatax[#1]#2{%
\if@filesw\immediate\write\@auxout{\string\citation{#2}}\fi%
  \def\@citea{}\@cite{\@for\@citeb:=#2\do%
    {\@citea\def\@citea{, }\@ifundefined% by Young
       {b@\@citeb}{{\bf ?}%
       \@warning{Citation `\@citeb' on page \thepage \space undefined}}%
{\csname b@\@citeb\endcsname}}}{#1}}%

\def\@citex[#1]#2{%
\if@filesw\immediate\write\@auxout{\string\citation{#2}}\fi%
  \def\@citea{}\@cite{\@for\@citeb:=#2\do%
    {\@citea\def\@citea{; }\@ifundefined% by Young
       {b@\@citeb}{{\bf ?}%
       \@warning{Citation `\@citeb' on page \thepage \space undefined}}%
{\csname b@\@citeb\endcsname}}}{#1}}%

\def\@biblabel#1{}

\newlength{\bibhang}
\makeatother

\begin{document}
 
\title{HST and Palomar Imaging of GRB 990123: Implications
for the Nature of Gamma-Ray Bursts and their Hosts}
%\title{Deep Imaging of the GRB~970228 Field with HST-STIS}
%\title{HST/STIS Observations of the Optical Counterpart to GRB~970228}

\author{Andrew S. Fruchter$^1$,
S. E. Thorsett$^{2,}$\altaffilmark{12},
Mark R. Metzger$^{3}$,
Kailash C. Sahu$^1$,
Larry Petro$^1$,
Mario Livio$^1$,
Henry Ferguson$^{1}$,
Elena Pian$^4$,
David W. Hogg$^{5,}$\altaffilmark{13},
Titus Galama$^6$,
Theodore R. Gull$^7$,
Chryssa Kouveliotou$^8$,
Duccio Macchetto$^{1,9}$,
Jan van Paradijs$^{6,14}$,
Holger Pedersen$^{10}$,
Alain Smette$^{7,11}$}

\affil{$^{1}$Space Telescope Science Institute, 3700 San Martin 
Drive, Baltimore, MD 21218, USA\\
$^{2}$Joseph Henry Laboratories and Dept.\ of Physics, Princeton 
University, Princeton, NJ 08544, USA\\
$^{3}$Department of Astronomy, Caltech, MS 105-24, Pasadena, CA 91125\\
$^{4}$Istituto di Tecnologie e Studio delle Radiazioni 
Extraterrestri, C.N.R., Via Gobetti 101, I-40129 Bologna, Italy \\
$^5$Institute for Advanced Study, Princeton, NJ 08540 \\
$^{6}$Astronomical Institute ``Anton Pannekoek'', University
of Amsterdam, Kruislaan 403, 1098 SJ Amsterdam, The Netherlands\\
$^7$Code 681, Laboratory for Astronomy and Solar Physics, 
Goddard Space Flight Center, Greenbelt, MD 20771, USA\\
$^8$NASA Marshall Space Flight Center, ES-84, Huntsville, AL 35812, USA\\
$^{9}$Affiliated to the Astrophysics Division, Space Science Department, 
European Space Agency\\
$^{10}$Copenhagen University Observatory, Juliane Maries Vej 30, D-2100, 
Copenhagen \"{A}, Denmark \\
$^{11}$National Optical Astronomy Observatories, P.O. Box 26732, 950 North Cherry
Avenue, Tucson, AZ 85726-6732, USA\\
$^14$Department of Physics, University of Alabama in Huntsville, Huntsville, AL
35899}

\altaffiltext{12}{Alfred P. Sloan Research Fellow}
\altaffiltext{13}{Hubble Fellow}

\begin{abstract}
We report on  HST and Palomar optical images of the field of
GRB~990123, obtained on 8 and 9 February 1999.  We find that the
optical transient (OT) associated with GRB~990123 is located on an
irregular galaxy, with magnitude $\rm{V}=24.20 \pm 0.15$.  The strong
metal absorption lines seen in the spectrum of the OT, along with the
low probability of a chance superposition, lead us to conclude that
this galaxy is the host of the GRB.  The OT is projected within the
$\sim1''$ visible stellar field of the host, nearer the edge than
the center.  We cannot, on this basis, rule out the galactic nucleus
as the site of the GRB, since the unusual morphology of the host may
be the result of an ongoing galactic merger, but our demonstration
that this host galaxy has extremely blue optical to infrared colors
more strongly supports an association between GRBs and star
formation.  We find that the OT magnitude on 1999 Feb 9.05, $\rm{V} =
25.45 \pm 0.15$, is about $1.5$\,mag fainter than expected from
extrapolation of the decay rate found in earlier observations.  A
detailed analysis of the OT light curve suggests that its fading has
gone through three distinct phases: an early rapid decline ($f_{\nu}
\propto t^{-1.6}$ for $t < 0.1$ days), a slower intermediate decline
power-law decay ($f_{\nu} \propto t^{-1.1}$ for $0.1 < t < 2$ days),
and then a more rapid decay (at least as steep as $f_{\nu} \propto
t^{-1.8}$ for $t > 2$ days).  The break to steeper slope at late
times may provide evidence that the optical emission from this GRB
was highly beamed.

%If this interpretation is
 % correct, the energy requirement for the GRB event could be reduced
  %from $\gsim 10^{54}$ to $\sim10^{52}$\,ergs.

\end{abstract}

\section*{Introduction}

The gamma-ray burst GRB~990123 was an astrophysical event of
astonishing proportions.  It was the brightest burst yet detected
with the wide-field cameras on the {\it BeppoSAX} satellite
\cite{fpf+99}, and has a total fluence among the brightest 0.3\% of
bursts detected by the BATSE instrument on the {\it Compton Gamma-Ray
Observatory} (CGRO).  Optical observations, which began with the ROTSE-I
telephoto array while the gamma-ray event was still in progress,
detected an optical transient (OT) that reached a peak magnitude
$\rm{V} \sim 9 $ about 40 seconds after the start of the burst
\cite{am99}.  Within hours, spectroscopy revealed metal absorption
lines in the spectrum of the OT at $z = 1.60$ \cite{kif+99,hac+99},
constraining the GRB redshift to be at least this great.  If the
gamma-ray burst emission was directed isotropically, the implied
energy release is $\gsim 2 \times 10^{54}$\,ergs.

Recognizing the importance of rapid observations of this object and
the extraordinary level of interest in the astronomical community,
{\it Hubble Space Telescope} (HST) observations were scheduled using
director's discretionary time, for immediate public release
\cite{bec99a,bec99b}.  To help ensure rapid access, we processed the
data on the day of the observation and made the resulting images and a
first report on the results of the imaging freely available on the web
\cite{fsf+99}.  In this paper we present a more detailed analysis, and
combine the HST data with further optical images we have obtained at
the Palomar 5-m telescope.  Together these observations allow us to
study both the light curve of the OT and the nature of the host galaxy
of GRB~990123.

\section*{The Observations}

The field of GRB~990123 was observed by HST over the course of three
orbits between 8 February 23:06:54 UT and 9 February 03:21:43 UT,
using STIS in clear aperture mode.  Two images of 650\,s each were
taken at each of six dither positions for a total exposure time of
7800\,s.  The images were processed using the standard STIS data
pipeline; however, a ``dark'' was first constructed using data
obtained the week before the observations, and this was used instead
of the standard calibration file, dramatically reducing the number of
hot pixels in the pipeline-calibrated image.  The standard pipeline
performed a first pass cosmic-ray removal using the two images at each
dither position.  Remaining cosmic rays and hot pixels were eliminated
and the images combined using the Drizzle algorithm and associated
techniques \cite{fh99}.  The final output image was created with an
output pixel size of $0\farcs025$ on a side, or one-half that of the
original pixels.  A ``pixfrac'' of 0.6 was used.  Figure~1 shows
the central region of the final image.

The total counts associated with the OT were determined by two methods:
1) obtaining the counts above
an estimated galactic background in an aperture with radius 4 output
pixels, and then applying an aperture correction (a factor of 1.5)
determined using a STIS PSF star obtained from the HST archives,
and 2) directly subtracting the PSF star from the image while scaling
to minimize the image residuals.  The values obtained by these methods agreed
to 5\%.
The
total counts in a box $2\farcs5$ on a side were then found, and the
counts from the OT were subtracted to obtain the counts associated with the
galaxy.

Photometric calibration of the images was performed using the
synthetic photometry package SYNPHOT in IRAF/STSDAS. 
We refer the reader to Fruchter \etal (1999c)
\nocite{fpt+99}
for further discussion of the STIS/CCD and its photometric 
calibration.
We find that on 1999 Feb 9.05 UT, the OT
of GRB~990123 had a magnitude of $\rm{V} = 25.45 \pm 0.15$, and the
galaxy on which it is superposed was $\rm{V} = 24.20 \pm 0.15$.
(Unless otherwise stated, all magnitudes in this paper are Vega
magnitudes, Landolt \nocite{lan92} 1992).

Observations of the GRB host were also obtained using the Palomar 5-m
telescope with the COSMIC camera in direct imaging mode on 1999 Feb
8.4-8.5 UT (approximately 12 h before the HST observations). The
detector in COSMIC is a 2048x2048 CCD with 24\,micron pixels,
projecting to $0.28"$ on the sky.  Three dithered 300\,s exposures
were obtained in B-band, 3x300\,s in V, and 4x300\,s in R.  
The seeing measured in the final images varies
from $1.0"$ in V to $1.5"$ in B.  
%The
%moon was fairly bright in these frames, increasing the background
%significantly. 

Photometric calibration was performed using images of the PG 1633
field (Landolt 1992).  A comparison to the fainter stars measured by
Nilakshi \etal (1999)\nocite{nmps99} shows agreement to better than
0.1\,mag.  Aperture photometry of the host galaxy was performed using
a 5-pixel radius aperture, and including a rough correction to large
aperture derived from the curve-of-growth measured
for bright nearby point sources.
%Since the galaxy itself is extended, this represents a slight
%underestimate of the total brightness, but given the seeing and
%aperture size this systematic offset should not exceed the statistical
%error.  
The resulting magnitudes are $\rm{B} = 24.4 \pm 0.2$, $\rm{V}
= 23.96 \pm 0.05$, and $\rm{R} = 23.62 \pm 0.05$.  Given the
magnitude measured for the OT in the HST images taken hours later,
we estimate that the galaxy alone is
approximately 0.3 mags fainter.

The USNO A2.0 stars surrounding GRB~990123 have been used by a
number of observers as standards for performing relative photometry on
the OT.  We have therefore used our observations to recalibrate 47 of
the USNO A2.0 standards that fall in the frame and are
sufficiently faint to avoid saturation.  We find that in this region,
the USNO calibration is $\sim0.2$\,mag too bright. 
This correction is in the opposite
sense of that reported by \nocite{ski99} Skiff (1999).

\section*{The Light Curve of the Optical Transient}

In Figure~2, we show the R band light curve of the counterpart of
GRB~990123, combining our observations with those reported in the
literature.  We have attempted to reduce all available observations to
the photometric standards measured by Nilakshi \etal
(1999)\nocite{nmps99}.  Observations by
Sagar \etal (1999) \nocite{spm+99} and Veillet (1999) \nocite{vei99}
have been directly referenced to these standards, and we have been
able to reference published photometry from 
Sokolov \etal (1999) \nocite{szn+99}, 
Garnavich \etal (1999), \nocite{gjsg99} 
Yadigaroglu \etal (1999), Yadigaroglu and Halpern (1999)
and Halpern \etal (1999)
\nocite{yhuk99,yh99,hylk99} to the same standards. Other
observations, including 
Zhu and Zhang (1999), Zhu, Chen and Zhang (1999),
\nocite{zz99,zcz99} and 
Masetti \etal (1999) \nocite{mpp+99} were made
relative to USNO A1.0 stars.
Observations made relative to USNO A2.0 stars 
\cite{ol99,lg99} 
were adjusted by 0.2\,mag, as indicated by our photometry of A2.0
stars described above.  Gunn r band observations made with the Palomar
1.5-m telescope \cite{gob+99} were reduced to Cousins R assuming
$\mbox{r}-\mbox{R}=0.45$. Observations using the ROTSE-I telephoto
array and an unfiltered CCD \cite{am99} were reported as approximate
V magnitudes relative to catalog values for nearby reference stars.
For these data, as well as the STIS data, we have estimated the R band
flux assuming $\mbox{V}-\mbox{R}\approx0.2$, consistent with the
measured colors of the transient at later times \cite{mpp+99}.

%Fireball models (see Piran 1999\nocite{pir99} for a recent review)
%predict a power-law decay of the OT, $t^{-\beta}$, though the
%power-law index $\beta$ may change with time (e.g., when the remnant
%goes from radiative to adiabatic evolution). The power-law slopes, and 
%the times of power-law breaks, can be directly compared with
%theoretical models.  
We have analyzed the OT light curve in
terms of a broken power law.  As seen in Fig.\,2, at least three
power-law segments are required to fully characterize the data.

First, it is important to note that the measured power-law index at
early times depends strongly on the assumed time origin.  One common
choice is the BATSE trigger time. However, this time is strongly
dependent on the particular design of the BATSE instrument, and
particularly on the trigger energy band, since both the gamma-ray
onset and duration were strongly energy dependent.  The OSSE and
COMPTEL instruments on CGRO detected a much narrower burst from
GRB~990123 at MeV energies \cite{msmk99,ckbp99} than did BATSE in the
hard X-ray.  In Fig.\,2 we have chosen the time origin to be the
time of peak hard X-ray and gamma-ray emission, which was independent
of energy.  With this choice of origin, the ROTSE data are
well-described by a simple power-law decay, with $\beta=1.6$.  Using
the BATSE trigger as an origin would steepen the curve at early times,
and produce a poorer fit.

Independent of the choice of time origin, R band observations made
between 0.16 and 2.75\,days after the GRB are well described by a
power law with index $\beta=1.09\pm0.05$. 
%similar to observed for the
%optical transients of GRB~970228 ($\beta=1.14\pm0.05$, Fruchter \etal
%1999b\nocite{fpt+99}) and GRB~990508 ($\beta=1.141\pm0.014$, Galama
%\etal 1998\nocite{ggv+98c}).  
The data are consistent with a break to
this shallower decay slope $\sim15$\,minutes after the GRB, though
more complicated light curve behavior during the data gap between
10\,min and 3.8\,hrs cannot be ruled out. Beginning 
one week after the GRB \cite{yhuk99}, 
the data are consistently fainter
than the extrapolation of the $\beta=1.09$ power law; the OT magnitude
measured by STIS is $\sim1.5$\,mags fainter than expected.  If the
fading behavior of the transient since day four is described by
another power law, its slope must be at least as steep as $\beta=1.8$,
as can be seen in Fig.~2, but a slope as steep as $\beta=2.5$ cannot
be ruled out.  Determining whether this is a true break in the
long-term fading behavior, or a temporary fluctuation around a long
term $t^{-1.1}$ power law (as was seen in the GRB~970228 light curve,
Fruchter \etal 1999c\nocite{fpt+99}) will require further observations.
If the OT continues to fade at $\beta=1.8$, it will become essentially
undetectable, even from HST, in the spring of 1999.

There are numerous possible physical causes for the observed breaks in
the OT light curve, and in the absence of broadband spectral
information a definitive classification of temporal breaks is
impossible.  One physically plausible way to account for the early
break from steep decay to shallower decay at early times is to suppose
that the early ($\lsim10^{-2}$\,days), rapidly fading optical flash
arises in the reverse shock that propagates from the fireball-ISM
boundary back into the ejecta material, while the more slowly fading
optical emission seen on timescales of a few hours arises in the
forward shock propagating into the ISM \cite{sp99b,sp99a}.  A
possibility for a break to a steeper power law at late times is the
spread of the opening angle of a strongly beamed 
(or jet dominated) flow \cite{rho97b,pir99}.  This occurs when fireball
material has decelerated to $\Gamma\sim\theta^{-1}$, where $\theta$ is 
the opening angle of the beam.  
%However, M\'esz\'aros and Rees have
%optical emission seen on timescales of a few hours arises in the 
%optical emission seen on timescales of a few hours arises in the 
%Since (roughly) $\Gamma\sim10$ at a 
%few days, the implied opening angle of the beam is $\sim0.1$ and the
%resulting energy requirements for the GRB event $\sim10^{-2}$ what is
%inferred with the assumption of isotropy, with important implications
%for both the GRB energy source and event rate.
However, M\'esz\'aros and Rees (1999)
\nocite{mr99} have pointed out that a break may occur at this time
due to the observer beginning to see the edge of the jet.  They
predict an increase in the power-law index of $0.75$, which is
approximately equal to the lower-limit of the break observed here.

\section*{The Nature of the Host Galaxy}

%When the first optical transient associated with a GRB was discovered
%\cite{vgg+97}, a question of paramount importance was whether the GRB
%was a Galactic or extragalactic phenomenon.  One of the deciding
%factors in this debate was the discovery that the OT was located on a
%faint galaxy \cite{slp+97,fpt+99}.  
Eleven
OTs associated with GRBs have been reported, at least nine of which are well
established, and with at most one exception these appear to lie on host
galaxies \cite{hf99}.  Reasonable models predict that in about
20\% of cases the host galaxy will be fainter than $\rm{R} \sim 27$,
and thus too faint to be detected by the ground-based telescopes that
have been used \cite{hf99}.  
The association between GRBs and
galaxies is hence well established, leading to a reasonable
presumption that a galaxy coincident with a GRB is its host.  The
particular case of GRB~990123 (as in 970508,\nocite{mdk+97} Metzger \etal 1997)
is even stronger, because spectra
obtained by the Keck and NOT telescopes show a deep metal line
absorption system at $z=1.60 $ \cite{kif+99,hap+99}, but
no evidence of other absorption line systems.

Given that the observed galaxy is at $z = 1.60$, what can we learn
about it from the optical observations presented here, and the K band
data reported by Bloom \etal 1999\nocite{bod+99}? (Note: Bloom
\etal also discuss the HST data analyzed here.)  In Fig.~3, we
compare the colors of the host galaxy with those of other objects in
the HDF-N \cite{wms96,dic97} and with those of two other host
galaxies, GRB~980703 \cite{bfk+98} and GRB~970228 \cite{cl98,fpt+99}.  As
is common in papers on the HDF, the colors are shown in AB magnitudes
(for all colors, AB mag $23.9 = 1\,\mu$Jy).  All three
hosts lie on the locus which defines the bluest edge of observed
galaxies.  Unfortunately, there are few galaxies with spectroscopic
redshifts $z \sim 1.6$ that are as faint as the host of GRB~990123;
however, one can compare this object to galaxies in the HDF which have
estimated spectroscopic redshifts of $z \sim 1.6$ using the catalog of
Fernandez-Soto, Lanzetta and Yahil (1999).  \nocite{fly99} One finds
that the host is among the bluest of galaxies at that redshift in
$\rm{V} - \rm{K}$, but it is not particularly bright.  Indeed there
are of order a dozen galaxies in that catalog in the range $1.3 < z <
1.9$ which are brighter in the {\it blue} (rest frame UV) than the
host of GRB~990123.  Therefore while the host is rapidly star-forming
for its mass (or population of old stars) it is not a particularly
bright galaxy.  

\section*{GRB Progenitors}

The position of the GRB on the host can also provide information on
the progenitors of these extraordinary outbursts.  We now have four
GRBs with HST imaging: 970228 \cite{slp+97,fpt+99}, 970508 \cite{fp98},
971214 \cite{odk+98} and 990123.  In all four cases, the OT occurs
superposed on the stellar field.  Thus neutron-star--neutron-star
mergers, which on occasion should happen well outside the
stellar field due to kicks given by supernovae at their birth
\cite{dc87,bsp98}, are perhaps disfavored both by this evidence, and by
the fact stated earlier that nearly all OTs have host galaxies.
However, it is possible that without a dense external working surface
no OT would appear \cite{mr93,sp97}, so by only localizing GRBs with
bright OTs we may be selecting for events which occur in stellar
fields.

If all GRBs at cosmological distances are produced by the same
mechanism, then the images of GRB~970228 
would have ruled out AGN as the source of GRBs: GRB~970228 lies at the
edge of an undistinguished galactic disk.  On the other hand it is
difficult to use GRB~990123 as further evidence against AGN, for if
one were to ask what local galaxy the host of GRB~990123 resembles,
one might well choose NGC~4038/4039---the ``Antennae,'' the most
well-known galaxy merger \cite{wsl+95}.  In the early stages of
merger, the massive black hole(s) need not be near the apparent center
of the remnant.  
Furthermore, if more than one mechanism causes cosmological GRBs,
then one must wonder whether the remarkable $0\farcs01$ coincidence of
GRB~970508 with the center of that regular host galaxy \cite{fpg+99}
suggests the presence of a nuclear starburst or a massive black hole.

\section*{Acknowledgements} 
We wish to thank Steven Beckwith, the Director of
STScI, for using Director's Discretionary Time to observe GRB~990123
and for making the data public.  We also thank Jen Christensen for assistance in
creating appropriate STIS dark files.
%Support for DWH was provided by
%Hubble Fellowship grant HF-01093.01-97A from STScI, which is operated
%by AURA under NASA contract NAS~5-26555.

Note: The reduced HST images discussed in this paper can be retrieved
in FITS format from http://www.stsci.edu/~fruchter/GRB/990123.
The GCN circulars mentioned in the bibliography are available from 
http://gcn.gsfc.nasa.gov.  

%\bibliographystyle{apj}
%\bibliography{journals_apj,grbrefs,new}

\begin{thebibliography}{}

\bibitem[\protect\citeauthoryear{Akerlof \& McKay}{Akerlof \&
  McKay}{1999}]{am99}
Akerlof, C.~W.,  \& McKay, T.~A. 1999.
\newblock IAU circular 7100

\bibitem[\protect\citeauthoryear{Beckwith}{Beckwith}{1999a}]{bec99b}
Beckwith, S. 1999a, GCN notice 254

\bibitem[\protect\citeauthoryear{Beckwith}{Beckwith}{1999b}]{bec99a}
Beckwith, S. 1999b, GCN notice 245

\bibitem[\protect\citeauthoryear{Bloom et~al.}{Bloom et~al.}{1998}]{bfk+98}
Bloom, J.~S.,  et~al. 1998, ApJ, 508, L21

\bibitem[\protect\citeauthoryear{Bloom et~al.}{Bloom et~al.}{1999}]{bod+99}
Bloom, J.~S., et~al. 1999, ApJ, submitted, astro-ph/9902182

\bibitem[\protect\citeauthoryear{Bloom, Sigurdsson, \& Pols}{Bloom
  et~al.}{1998}]{bsp98}
Bloom, J.~S., Sigurdsson, S.,  \& Pols, O.~R. 1998, MNRAS, submitted

\bibitem[\protect\citeauthoryear{Castander \& Lamb}{Castander \&
  Lamb}{1998}]{cl98}
Castander, F.~J.,  \& Lamb, D.~Q. 1998, in Gamma Ray Bursts: The 4th Huntsville
  Symposium, ed. C.~A. Meegan, R.~D. Preece, \& T.~M. Koshut (AIP Conference
  Proceedings, 428), 520

\bibitem[\protect\citeauthoryear{Connors et~al.}{Connors et~al.}{1999}]{ckbp99}
Connors, A., Kippen, R.~M., Barthelmy, S.,  \& Butterworth, P. 1999, GCN notice
  230

\bibitem[\protect\citeauthoryear{Dewey \& Cordes}{Dewey \& Cordes}{1987}]{dc87}
Dewey, R.~J.,  \& Cordes, J.~M. 1987, ApJ, 321, 780

\bibitem[\protect\citeauthoryear{Dickinson}{Dickinson}{1997}]{dic97}
Dickinson, M.~E. 1997,
  http://www.stsci.edu/ftp/science/hdf/clearinghouse/irim/irim\_hdf.html

\bibitem[\protect\citeauthoryear{Fernandez-Soto, Lanzetta, \&
  Yahil}{Fernandez-Soto et~al.}{1999}]{fly99}
Fernandez-Soto, A., Lanzetta, K.,  \& Yahil, A. 1999, ApJ, in press,
  astro-ph/9809126

\bibitem[\protect\citeauthoryear{Feroci et~al.}{Feroci et~al.}{1999}]{fpf+99}
Feroci, M., Piro, L., Frontera, F., Torroni, V., Smith, M., Heise, J.,  \&
  in~'t Zand, J. 1999.
\newblock IAU circular 7095

\bibitem[\protect\citeauthoryear{Fruchter \& Pian}{Fruchter \&
  Pian}{1998}]{fp98}
Fruchter, A.,  \& Pian, E. 1998, GCN notice 151

\bibitem[\protect\citeauthoryear{Fruchter et~al.}{Fruchter
  et~al.}{1999a}]{fsf+99}
Fruchter, A., Sahu, K., Ferguson, H., Livio, M.,  \& Metzger, M. 1999a, GCN
  notice 255

\bibitem[\protect\citeauthoryear{Fruchter \& Hook}{Fruchter \&
  Hook}{1999}]{fh99}
Fruchter, A.~S.,  \& Hook, R.~N. 1999, PASP, submitted, astro-ph/9808087

\bibitem[\protect\citeauthoryear{Fruchter et~al.}{Fruchter
  et~al.}{1999b}]{fpg+99}
Fruchter, A.~S., et~al. 1999b, ApJ, submitted, astro-ph/9903236

\bibitem[\protect\citeauthoryear{Fruchter et~al.}{Fruchter
  et~al.}{1999c}]{fpt+99}
Fruchter, A.~S., et~al. 1999c, ApJ, in press, astro-ph/9807295

\bibitem[\protect\citeauthoryear{Gal et~al.}{Gal et~al.}{1999}]{gob+99}
Gal, R.~R., Odewahn, S.~C., Bloom, J.~S., Kulkarni, S.~R.,  \& Frail, D.~A.
  1999, GCN notice 207

\bibitem[\protect\citeauthoryear{Garnavich et~al.}{Garnavich
  et~al.}{1999}]{gjsg99}
Garnavich, P., Jha, S., Stanek, K.,  \& Garcia, M. 1999, GCN notice 215

\bibitem[\protect\citeauthoryear{Halpern et~al.}{Halpern et~al.}{1999}]{hylk99}
Halpern, J.~P., Yadigaroglu, Y., Leighly, K.~M.,  \& Kemp, J. 1999, GCN notice
  257

\bibitem[\protect\citeauthoryear{Hjorth et~al.}{Hjorth et~al.}{1999a}]{hac+99}
Hjorth, J., et~al. 1999a, GCN notice 219

\bibitem[\protect\citeauthoryear{Hjorth et~al.}{Hjorth et~al.}{1999b}]{hap+99}
Hjorth, J., Andersen, M.~I., Pedersen, H., Zapatero-Osorio, M., Perz, E.,  \&
  A.~J.~Castro\, T. 1999b, GCN notice 249

\bibitem[\protect\citeauthoryear{Hogg \& Fruchter}{Hogg \&
  Fruchter}{1998}]{hf99}
Hogg, D.~W.,  \& Fruchter, A.~S. 1998, ApJ, submitted, astro-ph/9807262

\bibitem[\protect\citeauthoryear{Kelson et~al.}{Kelson et~al.}{1999}]{kif+99}
Kelson, D.~D., Illingworth, G.~D., Franx, M., Magee, D.,  \& van Dokkum, P.~G.
  1999.
\newblock IAU circular 7096

\bibitem[\protect\citeauthoryear{Lachaume \& Guyon}{Lachaume \&
  Guyon}{1999}]{lg99}
Lachaume, R.,  \& Guyon, O. 1999.
\newblock IAU circular 7096

\bibitem[\protect\citeauthoryear{Landolt}{Landolt}{1992}]{lan92}
Landolt, A.~U. 1992, AJ, 104, 340

\bibitem[\protect\citeauthoryear{Masetti et~al.}{Masetti et~al.}{1999}]{mpp+99}
Masetti, N., et~al. 1999, GCN notice 233

\bibitem[\protect\citeauthoryear{Matz et~al.}{Matz et~al.}{1999}]{msmk99}
Matz, S.~M., Share, G.~H., Murphy, R.,  \& Kurfess, J.~D. 1999, GCN notice 231

\bibitem[\protect\citeauthoryear{M\'esz\'aros \& Rees}{M\'esz\'aros \&
  Rees}{1993}]{mr93}
M\'esz\'aros, P.,  \& Rees, M.~J. 1993, ApJ, 405, 278

\bibitem[\protect\citeauthoryear{M\'esz\'aros \& Rees}{M\'esz\'aros \&
  Rees}{1999}]{mr99}
M\'esz\'aros, P.,  \& Rees, M.~J. 1999, ApJ, submitted, astro-ph/9902367

\bibitem[\protect\citeauthoryear{Metzger et~al.}{Metzger et~al.}{1997}]{mdk+97}
Metzger, M.~R., Djorgovski, S.~G., Kulkarni, S.~R., Steidel, C.~C., Adelberger,
  K.~L., Frail, D.~A., Costa, E.,  \& Frontera, F. 1997, Nature, 387, 879

\bibitem[\protect\citeauthoryear{Nilakshi et~al.}{Nilakshi
  et~al.}{1999}]{nmps99}
Nilakshi, R. K. S.~Y., Mohan, V., Pandey, A.~K.,  \& Sagar, R. 1999, Bull.\
  Astr.\ Soc.\ India, in press

\bibitem[\protect\citeauthoryear{Odewahn et~al.}{Odewahn et~al.}{1998}]{odk+98}
Odewahn, S.~C.,  et~al. 1998, ApJ, 509, L5

\bibitem[\protect\citeauthoryear{Ofek \& Leibowitz}{Ofek \&
  Leibowitz}{1999}]{ol99}
Ofek, E.,  \& Leibowitz, E.~M. 1999.
\newblock IAU circular 7096

\bibitem[\protect\citeauthoryear{Piran}{Piran}{1999}]{pir99}
Piran, T. 1999, Phys. Rep., in press

\bibitem[\protect\citeauthoryear{Rhoads}{Rhoads}{1997}]{rho97b}
Rhoads, J.~E. 1997, ApJ, 487, L1

\bibitem[\protect\citeauthoryear{Sagar et~al.}{Sagar et~al.}{1999}]{spm+99}
Sagar, R., Pandey, A.~K., Mohan, V., Nilakshi, R. K. S.~Y., Bhattacharya, D.,
  \& Castro-Tirado, A.~J. 1999, accepted, astro-ph/9902196

\bibitem[\protect\citeauthoryear{Sahu et~al.}{Sahu et~al.}{1997}]{slp+97}
Sahu, K.~C., et~al. 1997, Nature, 387, 476

\bibitem[\protect\citeauthoryear{Sari \& Piran}{Sari \& Piran}{1997}]{sp97}
Sari, R.,  \& Piran, T. 1997, ApJ, 485, 270

\bibitem[\protect\citeauthoryear{Sari \& Piran}{Sari \& Piran}{1999a}]{sp99b}
Sari, R.,  \& Piran, T. 1999a, preprint, astro-ph/9902009

\bibitem[\protect\citeauthoryear{Sari \& Piran}{Sari \& Piran}{1999b}]{sp99a}
Sari, R.,  \& Piran, T. 1999b, preprint, astro-ph/9901338

\bibitem[\protect\citeauthoryear{Skiff}{Skiff}{1999}]{ski99}
Skiff, B.~A. 1999.
\newblock IAU circular 7098

\bibitem[\protect\citeauthoryear{Sokolov et~al.}{Sokolov et~al.}{1999}]{szn+99}
Sokolov, V., Zharikov, S., Nicastro, L., Feroci, M.,  \& Palazzi, E. 1999, GCN
  notice 209

\bibitem[\protect\citeauthoryear{Veillet}{Veillet}{1999}]{vei99}
Veillet, C. 1999, GCN notice 253

\bibitem[\protect\citeauthoryear{Whitmore et~al.}{Whitmore
  et~al.}{1995}]{wsl+95}
Whitmore, B.~C., Schweizer, F., Leitherer, C., Born, K.,  \& Robert, C. 1995,
  AJ, 106, 1354

\bibitem[\protect\citeauthoryear{{Williams} et~al.}{{Williams}
  et~al.}{1996}]{wms96}
{Williams}, R.~E., et~al. 1996, AJ, 112, 1335

\bibitem[\protect\citeauthoryear{Yadigaroglu \& Halpern}{Yadigaroglu \&
  Halpern}{1999}]{yh99}
Yadigaroglu, I.~A.,  \& Halpern, J.~P. 1999, GCN notice 248

\bibitem[\protect\citeauthoryear{Yadigaroglu et~al.}{Yadigaroglu
  et~al.}{1999}]{yhuk99}
Yadigaroglu, I.~A., Halpern, J.~P., Uglesich, R.,  \& Kemp, J. 1999, GCN notice
  242

\bibitem[\protect\citeauthoryear{Zhu, Chen, \& Zhang}{Zhu et~al.}{1999}]{zcz99}
Zhu, J., Chen, J.~S.,  \& Zhang, H.~T. 1999, GCN notice 217

\bibitem[\protect\citeauthoryear{Zhu \& Zhang}{Zhu \& Zhang}{1999}]{zz99}
Zhu, J.,  \& Zhang, H.~T. 1999.
\newblock IAU circular 7095

\end{thebibliography}

\eject
\section*{Figure Captions}

Figure~1:  The central $3.''2$ 
of the HST image of the field of GRB~990123.  
The OT is the bright point source at the center
of the image.  North is up, East to the left.   
The two small objects to the south-east have been included
in the measured magnitude of the galaxy because
their projected separation from the main body of the galaxy,
$< 7$ kpc, makes it likely that they are now
or will soon be part of the host galaxy.

Figure~2: The R band light curve of the OT associated with GRB~990123.
All points, except for the HST point (solid fill, on right), were
taken from the literature as discussed in the text and reduced to a
common flux standard with the galaxy flux subtracted.  Error
bars are shown where available ($1\sigma$), arrows indicate 95\%
confidence upper limits. 
%The light curve appears to have three well
%defined sections: a rapid power-law decline in the first ten minutes, a
%lower power-law decay over the next two days, and then another rapid
%decay (details of the power-law fits are given in the text).  The
%latter rapid decay, if continued unbroken, could make the OT
%essentially unobservable, even by HST, within 3 months after outburst.

Figure~3:   The host galaxies of
GRB 970228, 980703 and 990123 on a color-magnitude
diagram along with objects in the HDF.  Stars are displayed as light blue stars.
Galaxies in the HDF with spectrophotometric redshift are shown
in green for $0 < z < 1$, red for $1 < z < 2$ and magenta for
$z > 2$.  The GRB host galaxies are shown in dark blue.
Upper and lower limits are shown with arrows.  Because
we do not have an H magnitude for the host of GRB~990123, its K magnitude
has been used instead.  We would expect measured H values
to be slightly brighter and bluer than shown, which would move the
point down and to the left.  From
brighter to fainter mag (left to right) the GRB hosts
are 980703, 990123, 970228.

\eject

\begin{figure}
\centerline{\epsfig{file=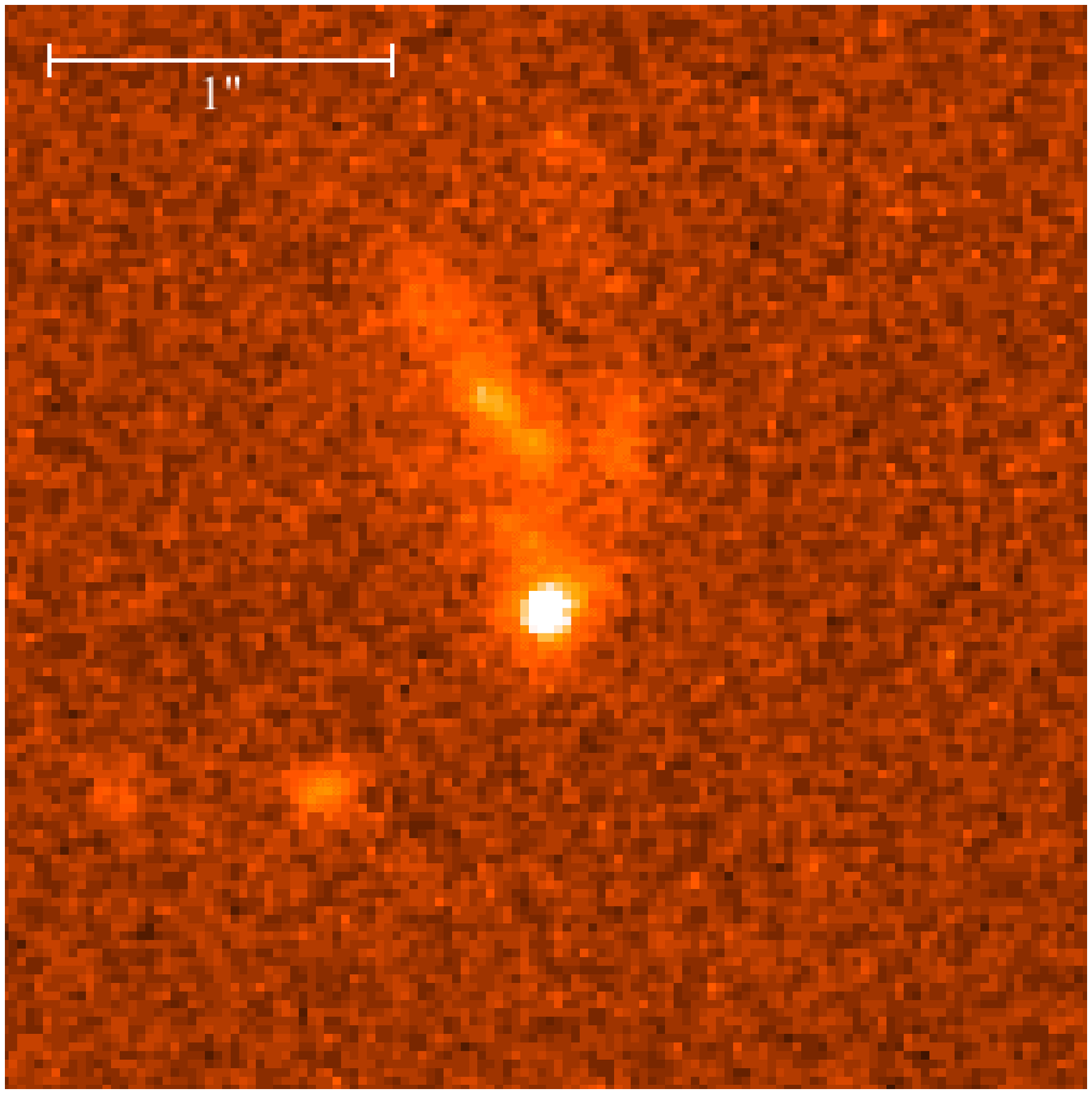}}
\end{figure}

\eject
\begin{figure}
\centerline{\epsfig{file=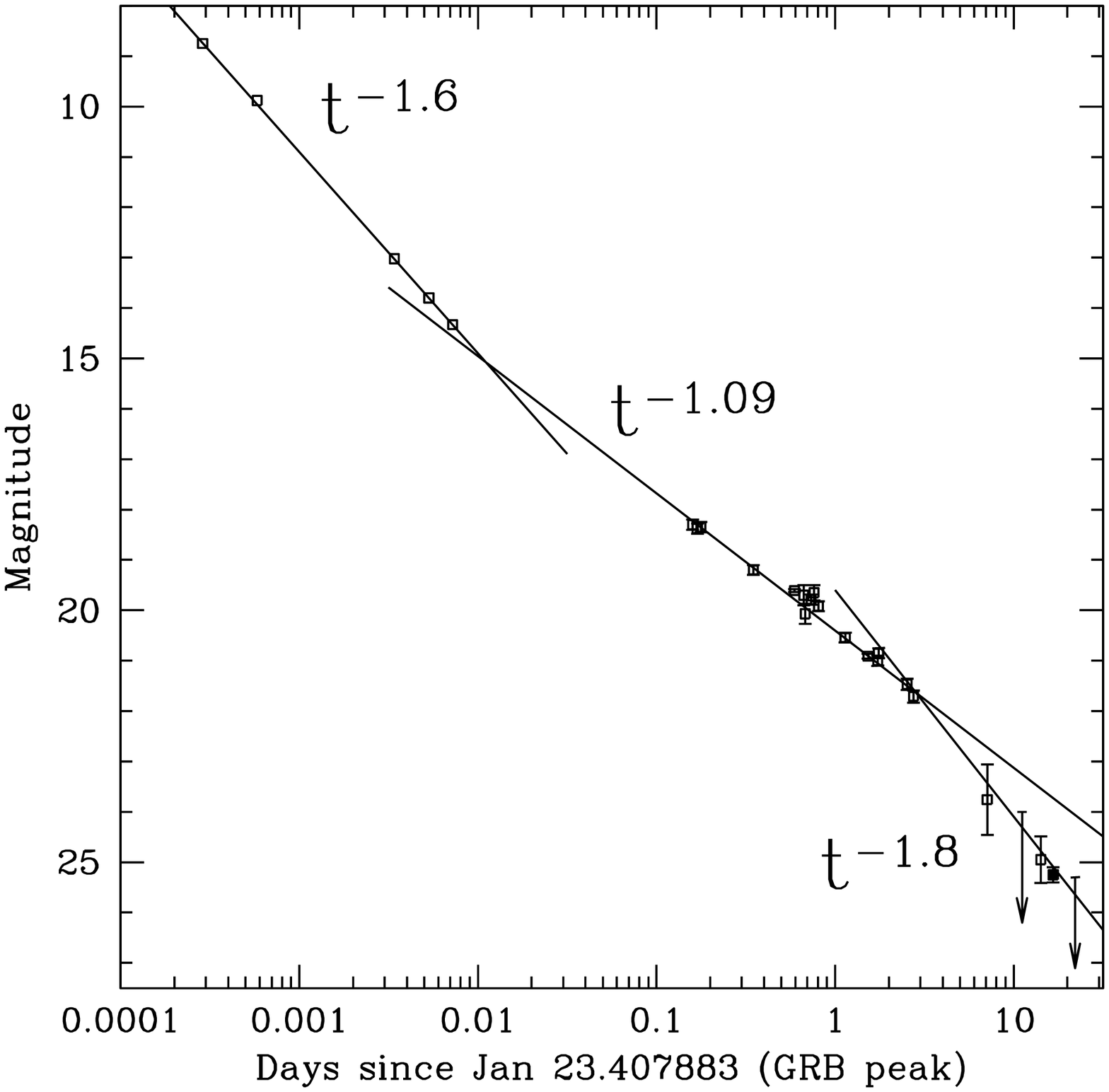}}
\end{figure}

\eject
\begin{figure}
\centerline{\epsfig{file=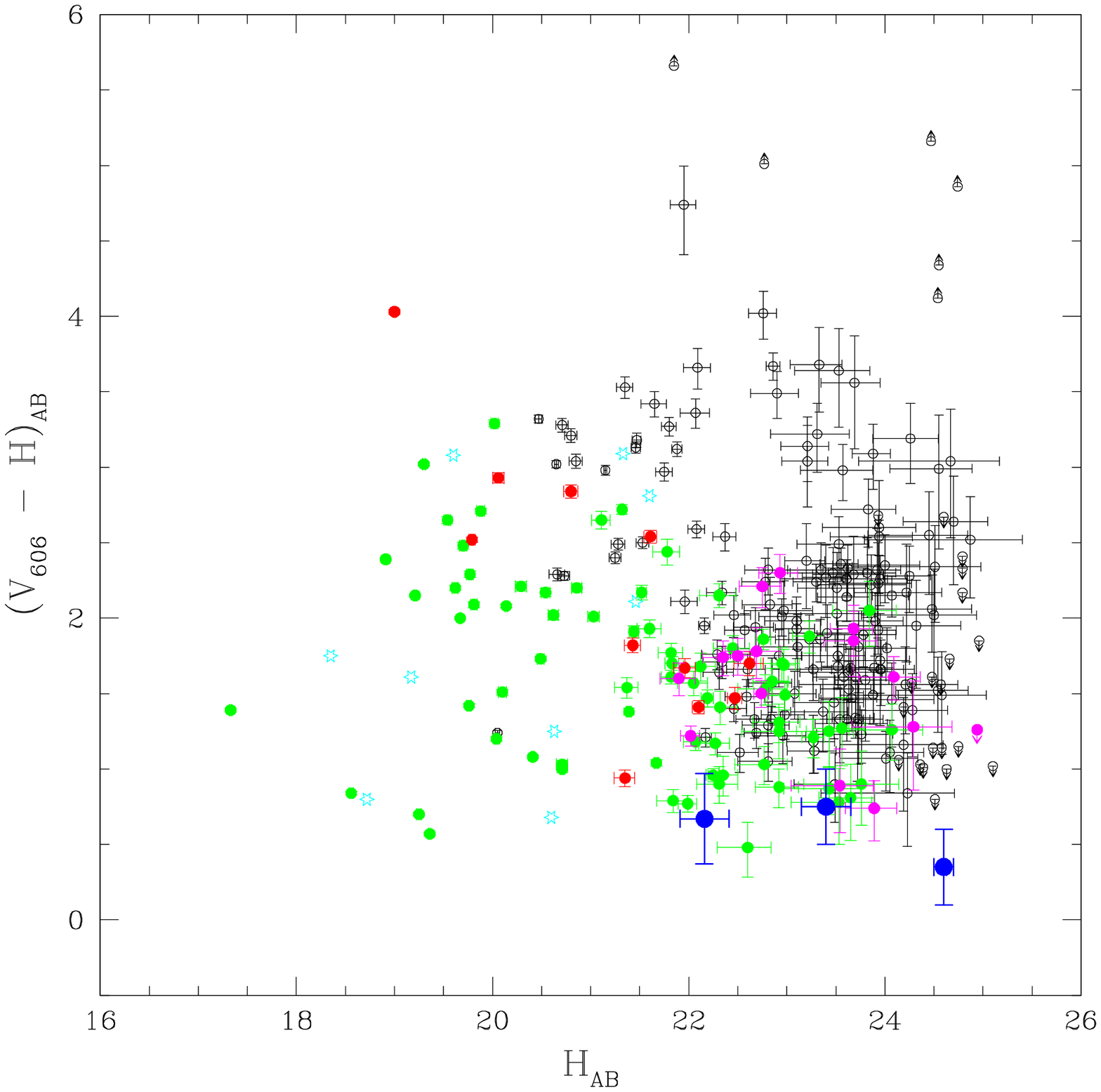}}
\end{figure}
\end{document}